**Demonstration of a Tuneable Coupler for Superconducting Qubits Using Coherent, Time Domain, Two-Qubit Operations**

R. C. Bialczak[1], M. Ansmann[1], M. Hofheinz[1], M. Lenander[1], E. Lucero[1], M. Neeley[1], A. D. O'Connell[1], D. Sank[1], H. Wang[1], M. Weides[1], J. Wenner[1], T. Yamamoto[1,2], A. N. Cleland[1], and J. M. Martinis[1*]

[1]*Department of Physics, University of California, Santa Barbara, CA 93106, USA*

[2] *Green Innovation Research Laboratories, NEC Corporation, Tsukuba, Ibaraki 305-8501, Japan*

*e-mail: martinis@physics.ucsb.edu

**Researchers using superconducting qubits have recently made significant progress towards implementing a quantum computer (1-11). However, decoherence and scalability continue to limit performance and must be addressed (12). It is now apparent that tuneable couplers (13,14) will be an important building block for scaling up superconducting qubits because they enable the interactions between qubits to be simply and directly controlled while avoiding frequency-crowding and residual coupling issues inherent to traditional frequency-tuned architectures (15). Tuneable couplers have been demonstrated using superconducting qubits (13, 16-18). However, these designs required the coupled/coupling elements to be in close proximity and showed either limited time-domain control or low on/off ratios making them difficult to scale up and of limited use in realistic qubit operations. We present a design which, for the first time in a single device, avoids these limitations. We experimentally show that using this tunable coupler, the inter-qubit coupling strength can be tuned with nanosecond time resolution and large on/off ratio within a sequence of qubit operations that**

**mimics actual use in an algorithm. The design is also self-contained and physically separate from the qubits, allowing the coupler to couple elements over long distances, a feature that will most likely be essential for a scalable quantum computer. This also allows the coupler to be used to couple other superconducting circuit elements in addition to qubits.**

The electrical circuit for the coupled Josephson phase qubits (5,7,19) is shown in Figure 1a and a corresponding optical micrograph is shown in Figures 1b-f. Each phase qubit (Figures 1b-c) is a non-linear resonator built from a Josephson junction, and external shunting capacitors and inductors, $C = 1\,\text{pF}$ and $L = 750\,\text{pH}$. When biased close to the critical current, the junction and its parallel loop inductance $L$ give rise to a cubic potential whose energy eigenstates are unequally spaced. The two lowest levels are used as the qubit states $|0\rangle$ and $|1\rangle$, with transition frequency $\omega_{10}/2\pi$. Logic operations (X and Y rotations) are performed by applying microwave pulses $I_{\mu W}^{A,B}$ at this transition frequency, whereas changes in frequency (Z rotations) and measurement are produced by ~ns pulses in the bias current $I_Z^{A,B}$ (5,7).

The tuneable coupling element (Figure 1d-f) is a three-terminal device constructed using a fixed negative mutual inductance $-M$ (Figure 1e) and a single, current-biased Josephson junction (Figure 1f) that acts as a tuneable positive inductance $L_c$. This inductance changes with coupler bias $I_{cb}$ according to (20)

$$L_c = \frac{\Phi_0}{2\pi I_{c0}\sqrt{1-(I_{cb}/I_{c0})^2}} \qquad (1)$$

where $I_{c0} \approx 1.58\,\mu\text{A}$ is the critical current. The interaction Hamiltonian between qubits A and B for the tuneable coupler is (14)

$$H_{int} \approx \frac{\hbar\Omega_c}{2}\left(\sigma_{x,A}\sigma_{x,B} + \frac{1}{6\sqrt{N_A N_B}}\sigma_{z,A}\sigma_{z,B}\right) \qquad (2)$$

with

$$\Omega_c = \frac{M - L_c}{(L_M + L_s)^2 \omega_{10} C} \tag{3}$$

where $\sigma$ is the Pauli operator, $N$ is the normalized well depth, and $\Omega_c \propto M - L_c$ is the adjustable coupling strength (13).

The direct connection of the qubits through this circuit allows for strong coupling. To reduce the coupling magnitude to the desired 50 MHz range, series inductors $L_s \approx 2657$ pH (Figure 1d) are inserted in the connecting wires that are significantly larger than for the mutual inductance element $L_M \approx 390$ pH and $M \approx 190$ pH. Because the number of levels in the potentials of qubits A and B are typically $N_A = N_B \approx 5$, the $\sigma_{z,A} \sigma_{z,B}$ term in equation *(2)* gives a small contribution of approximately 0.03 to the coupling strength. This interaction does not affect the results presented here and can be effectively removed using a simple refocusing sequence if needed. With parameters $I_{c0} \approx 1.58\,\mu A$ and $M=190$ pH, and full adjustment of the bias current, coupling strengths $\Omega_c / 2\pi$ were varied from approximately 0 MHz to 100 MHz. The values of $I_{c0}$ and $M$ can be chosen so that other ranges of coupling strength, both positive and negative in sign, are possible.

The direct-current connections between the coupler bias and the qubits produce small shifts in the qubit frequency due to changes in the coupler bias. These shifts can be readily compensated for using the qubit biases $I_Z^{A,B}$ and are discussed in the Supplementary Information, along with a coupler reset protocol (21).

The coupler can be operated in two modes. In the simplest "static" mode, the coupler is held at a fixed strength throughout a two-qubit pulse sequence. A more realistic "dynamic" mode uses a fast nanosecond-scale pulse to turn the coupler on and off during a control sequence that contains both single and two qubit operations and measurement.

Using the static mode, we performed spectroscopy at a fixed coupler bias, which allowed us to measure the energy splitting $\Omega_c/2\pi$ at the avoided level crossing where the detuning $\Delta/2\pi = f_{0A} - f_{0B}$ between the two qubits was zero. The pulse sequence for this experiment is shown in Figure 2a. A representative subset of crossings for several coupler biases are shown in Figures 2b-e. The coupler clearly modulates the size of the spectroscopic splitting, and allows the setting and measuring of the coupling strength $\Omega_c/2\pi$ (5,14,22). Although the data for 0 MHz shows no apparent splitting, the resolution at zero coupling is limited to $\pm 1.5$ MHz by the 3 MHz linewidth, which is slightly power-broadened. Sub-megahertz resolution of the minimum coupling strength is obtained from the time-domain experiments, as discussed below.

Since the fidelity of gate operations is limited by qubit coherence times (5), we want to reduce gate times by using strong coupling. For devices with fixed capacitive coupling, however, this strategy cannot be used effectively because of the rapid rise in measurement crosstalk with increased coupling (22,23). Therefore, it is important to demonstrate that measurement crosstalk can be reduced to a minimal value when the coupler is turned off. As shown is Figure 3a, we determine measurement crosstalk by driving only one qubit with Rabi oscillations, and then simultaneously measuring the excitation probabilities of both qubits (23). The undriven qubit ideally shows no response, with the ratio of the amplitudes of the oscillations giving a quantitative measure of the measurement crosstalk. The Rabi data is shown in Figures 3b and 3c for both the driven and undriven qubits, using representative coupling strengths of 0 MHz and ~17 MHz. The measurement crosstalk is plotted as a function of coupler bias in Figure 3d, where there is a broad region of coupler bias where the measurement crosstalk amplitude is minimized. This tuneable coupler allows operation of phase qubits at large coupling strengths without the drawback of large measurement crosstalk.

The "dynamic" mode of operation tests coupler performance with a sequence that mimics actual use in an algorithm. Turning off the coupler dynamically during single qubit operations prevents stray-coupling errors that normally would be present in an always-on architecture. As illustrated in Figure 4a, the coupling is first set to zero and a single qubit is excited to the $|1\rangle$ state. The qubits are then biased to within $\Delta/2\pi$ of resonance and the coupling is turned on to a coupling strength $\Omega_c/2\pi$ for an interaction time $t_{swap}$. Finally, the coupling is turned off and the qubit states are measured. The coupling produces a two-qubit swap operation which arises from the $\sigma_x\sigma_x$ operator in Eq. (2), which is the basis for universal gate operations (5). In Figure 4b, we show swap data for two representative settings for on and off coupling, $\Delta/2\pi \approx 0$ MHz at $\Omega_c/2\pi \approx 0$ MHz and $\Delta/2\pi \approx 0$ MHz at $\Omega_c/2\pi \approx 40$ MHz. In Figure 4c, swap data are shown where the detuning, $\Delta/2\pi$ was varied for several representative coupling strengths $\Omega_c/2\pi \approx 0, 11, 27, 45,$ and 100 MHz. The swaps exhibit the expected chevron pattern for the resonant interaction (22). In Figure 4d the swap frequency is plotted versus coupler bias for $\Delta/2\pi \approx 0$ MHz. Coupling strengths up to 100 MHz are possible, although we find that the decay times of the swaps degrade above 60 MHz, presumably due to the coupler bias approaching the critical current of the coupler junction.

The determination of the minimum coupling strength, which quantifies how well the interaction can be turned off, is limited by the minimum detectable swap frequency of the qubits. This frequency, in turn, is limited by qubit decoherence. A comparison of the off-coupling data in Figure 4b to simulations (see Supplementary Information) shows that the smallest resolvable coupling strength is no greater than 0.1 MHz. Given this upper-bound on the minimum coupling strength, the measured on/off ratio for the swap interaction is approximately 100 MHz/0.1 MHz = 1000.

Stray capacitances and inductances in the circuit introduce stray coupling that may limit the on/off ratio. In this design, the greatest contribution to stray coupling comes from the small, inherent capacitance of the coupler junction. The coupler junction has a self-resonance frequency of $\omega_{c0}/2\pi \sim 30$ GHz, which implies that its effective inductance $L_c[1-(\omega/\omega_{c0})^2]$ changes in value from $\omega/2\pi = 0$ and 6 GHz by ~4% (14). The $\sigma_z\sigma_z$ and $\sigma_x\sigma_x$ interactions in Eq. (2) will thus turn on and off at slightly different biases and, along with virtual transitions, will limit how far the coupler can be turned off (14, 15). A useful feature of this coupler is that this imperfection can be compensated for by including a small shunt capacitor across the mutual inductance (14), which should allow on/off ratios up to $10^4$.

The performance of this coupler, especially its ability to strongly couple qubits over long distances, makes it a promising "drop-in" module for scalable qubit architectures. The demonstration of coupling adjustment over nanosecond time scales, along with a large on/off ratio allows the implementation of algorithms that require on-the-fly tuning of the coupling strength (24). We also foresee the use of this coupling circuit in other applications, such as superconducting parametric amplifiers (25) or in coupling qubits to readout circuitry, superconducting resonators, or other circuit elements.

**Acknowledgements**

Devices were made at the UCSB Nanofabrication Facility, a part of the NSF-funded National Nanotechnology Infrastructure Network. We thank A. N. Korotkov, R. A. Pinto, and M. R. Geller for discussions. This work was supported by IARPA (grant W911NF-04-1-0204) and by the NSF (grant CCF-0507227). RCB designed and fabricated the samples, designed and performed experiments, and analysed the data. RCB co-wrote the paper with JMM and ANC who also supervised the project. MA provided assistance with device design and mask layout. All authors contributed to experiment set-up, sample design or sample fabrication. Correspondence and requests for materials should be addressed to JMM (martinis@physics.ucsb.edu).

# References

1. T. D. Ladd, F. Jelezko, R. Laflamme, Y. Nakamura, C. Monroe, J. L. O'Brien. Quantum computers. *Nature* **464**, 45-53 (4 March 2010).

2. Barbara Goss Levi. Superconducting qubit systems come of age. *Physics Today* **62**, 14 (2009).

3. Yamamoto, Y., Pashkin, Y. A., Astafiev, O., Nakamura, Y. & Tsai, J. S. Demonstration of conditional gate operation using superconducting charge qubits. *Nature* **425**, 941–944 (2003).

4. Plantenberg, J. H., de Groot, P. C., Harmans, C. J. & Mooij, J. E. Demonstration of controlled-NOT quantum gates on a pair of superconducting quantum bits. *Nature* **447**, 836–839 (2007).

5. R. C. Bialczak *et al*. Quantum process tomography of a universal entangling gate implemented with Josephson phase qubits. *Nature Physics* **6**, 409 - 413 (2010).

6. L. DiCarlo *et al*, Demonstration of two-qubit algorithms with a superconducting quantum processor. *Nature* **460**, 240 (2009).

7. M. Steffen *et al.* Measurement of the Entanglement of Two Superconducting Qubits via State Tomography. *Science* **313**, 1423 (2006).

8. M. Neely *et al.* Process tomography of quantum memory in a Josephson-phase qubit coupled to a two-level state. *Nature Physics* **4**, 523 (2008).

9. A. A. Houck *et al.* Generating single microwave photons in a circuit. *Nature* **449**, 328 (2007).

10. M. Hofheinz *et al.* Synthesizing arbitrary quantum states in a superconducting resonator. *Nature* **459**, 546 (2009).

11. A. D. O'Connell *et al.* Quantum ground state and single-phonon control of a mechanical resonator. *Nature* **464**, 697 (2010).

12. D. P. DiVincenzo. The Physical Implementation of Quantum Computation. *Fortschritte der Physik* **48**, 771 (2000).

13. T. Hime, *et al.* Solid-State Qubits with Current-Controlled Coupling. *Science* **314**, 1427 (2006).

14. R. A. Pinto, A. N. Korotkov, M. R. Geller, V. S. Shumeiko, J. M. Martinis. Analysis

of a tuneable coupler for superconducting phase qubits. *arXiv:1006.3351*

15. S. Ashhab *et al.* Interqubit coupling mediated by a high-excitation-energy quantum object. *Phys. Rev. B* **77**, 014510 (2008).

16. A. O. Niskanen *et al.* Quantum coherent tunable coupling of superconducting qubits. *Science* **316** (5825), 723-6 (2007).

17. M.S. Allman *et al.* rf-SQUID-Mediated Coherent Tunable Coupling between a Superconducting Phase Qubit and a Lumped-Element Resonator. *Phys. Rev. Lett.* **104**, 177004 (2010).

18. R. Harris *et al.* Sign- and Magnitude-Tunable Coupler for Superconducting Flux Qubits. *Phys. Rev. Lett.* **98** 177001 (2007).

19. J. M. Martinis. Superconducting Phase Qubits. *Quantum Information Processing* **8**, 81 (2009).

20. A. Barone, G. Paterno, *Physics and Applications of the Josephson Effect* (John Wiley & Sons, New York, NY 1982).

21. T. A. Palomaki *et al.* Initializing the flux state of multiwell inductively isolated Josephson junction qubits. *Phys. Rev B* **73**, 014520 (2006).

22. R. McDermott *et al.* Simultaneous State Measurement of Coupled Josephson Phase Qubits. *Science* **307** 1299 (2005).

23. M. Ansmann *et al.* Violation of Bell's inequality in Josephson phase qubits. *Nature* **461**, 504 (2009).

24. A. Galiautdinov and M. Geller. Controlled-NOT gate design for Josephson phase qubits with tunable inductive coupling: Weyl chamber steering and area theorem. *arXiv:quant-ph/0703208*.

25. N. Bergeal *et al.* Phase-preserving amplification near the quantum limit with a Josephson ring modulator. *Nature* **465**, 64 (2010).

**Figure 1. Device circuit and micrograph of two Josephson phase qubits with a tuneable coupler.** The two qubits are shown in red and blue in the circuit and in boxes *b* and *c* in the lower micrograph, and are described in detail in prior publications (19, 5, 7). The inductors $L_s$, $L_M$, and the mutual inductance $M$, which form the non-tuneable part of the coupler, are shown in purple and green and in boxes *d* and *e*. The current-biased coupler junction, which forms the tuneable element, is shown in orange and in box *g*. The inductor $L_z = 9$ nH isolates the coupler from the bias circuit. The entire coupler, a modular three-terminal device, is shown by the dashed box. The distances, *r*, can be made longer to couple elements over greater distances.

**Figure 2. Tuning the spectroscopic splitting. (a)** Pulse sequence for qubit spectroscopy. The coupler bias (green) is set to the value $I_{cb}$, and is kept at this constant value throughout this static experiment. The dashed line indicates the coupler bias level that corresponds to zero coupling strength, $\Omega_c / 2\pi \approx 0$ MHz. The qubits (red and blue) are initially detuned by 200 MHz and each starts in the $|0\rangle$ state. The bias of qubit B is then adjusted to set its transition frequency a distance $\Delta / 2\pi$ from the qubit A frequency, $f_{0A}$. A microwave pulse of frequency $f_{\mu w}$ and duration ~2 µs is then applied to each qubit. Both qubit states are then determined using a single-shot measurement. The panels **(b-e)** are plots of the measured probability $P_{10}$ (A excited) and $P_{01}$ (B excited) versus the detuning frequency $\Delta / 2\pi = f_{0A} - f_{0B}$ and the difference in microwave and qubit A frequencies $f_{\mu w} - f_{0A}$. Each panel shows a different coupler bias $I_{cb}$ that increases from left to right (top green arrow). The splitting size $\Omega_c / 2\pi$, measured as the minimum distance between the two resonance curves, shows the coupling strength being adjusted by the coupler. The dotted lines are a fit of the avoided level crossings.

**Figure 3. Turning off measurement crosstalk. (a)** Pulse sequence for determining the measurement crosstalk as a function of coupler bias. The coupler (green) is set to a static

bias $I_{cb}$. The coupler bias level corresponding to zero coupling strength, $\Omega_c/2\pi \approx 0$ MHz, is indicated by the dashed line. The qubits remain detuned by 200 MHz throughout the experiment, and only one qubit (shown here, A) is excited with microwaves. **(b)** For the tunable coupler turned off, we plot $P_{x1} = P_{01} + P_{11}$ and $P_{1x} = P_{10} + P_{11}$ versus the Rabi pulse time $t_{\text{rabi}}$. Rabi oscillations in $P_{1x}$ (qubit A) are observed, with only a small amplitude oscillation of $P_{x1}$ (qubit B) coming from measurement crosstalk. **(c)** Same as for (b), but with coupling turned on to 17 MHz. **(d)** Measurement crosstalk amplitude as a function of coupler bias $I_{cb}$ for the case of Rabi drive on qubit A (red) and qubit B (blue). For the case of drive on qubit A (qubit B), crosstalk amplitude is displayed as the ratio of the amplitudes of the oscillations of $P_{x1}$ ($P_{1x}$) to that of $P_{1x}$ ($P_{x1}$).

**Figure 4. Demonstration of dynamic coupler operation via swap experiment. (a)**
For the pulse sequence, the coupler (green) is first set to the coupler bias value corresponding to $\Omega_c/2\pi \approx 0$ MHz, as measured previously. The qubits (red and blue) are initially detuned by $\Delta/2\pi = 200$ MHz and start in the $|0\rangle$ state. A $\pi$ microwave pulse is then applied to qubit A, exciting it to the $|1\rangle$ state. The coupling interaction remains off during this pulse to minimize errors resulting from two-qubit interactions. A fast bias pulse then detunes qubit B from qubit A by a frequency $\Delta/2\pi$, and at the same time compensates for qubit bias shifts due to the coupler. Simultaneously, the coupler is turned on to a bias $I_{cb}$ using a fast bias pulse with ~2 ns rise and fall times. The coupler and qubit biases are held at these values for a time $t_{swap}$, allowing the two-qubit system to evolve under the swap operation. The qubits are then detuned again to $\Delta/2\pi = 200$ MHz and $I_{cb}$ is set back to zero coupling strength, allowing for a crosstalk-free single-shot measurement of the two-qubit probabilities $P_{AB} = \{P_{01}, P_{10}, P_{11}\}$. **(b)** The measured two-qubit probabilities $P_{01}$, $P_{10}$, and $P_{11}$ are plotted versus $t_{swap}$ for qubits on resonance $\Delta/2\pi = 0$ and two sets of coupling $\Omega_c/2\pi$, corresponding to off (top) and on (bottom) conditions. **(c)** Measured qubit probabilities $P_{01}$ (red, top panel) and $P_{10}$ (blue, bottom panel) are plotted versus $t_{swap}$ and qubit detuning $\Delta/2\pi$ for representative coupling strengths $\Omega_c/2\pi$ = ~0 MHz, 11 MHz, 27 MHz, 45 MHz, and 100 MHz. **(d)** Swap frequency $\Omega_c/2\pi$ versus coupler bias $I_{cb}$ for all coupler biases measured in this experiment (solid dots). Solid red line is theory obtained from Eq. (3) and measured device parameters.

## Supplementary Information

**Coupler reset protocol**

A direct-current connection between the coupler and qubits requires that the coupler must be reset. Because the Josephson inductance of the qubit junctions is much larger than the shunt inductor $L$, current flowing from the coupler mostly flows through $L$. As a result, the coupler junction is effectively shunted by two loops with net inductance $(L_M + L_s + L)/2$, as shown in Supplementary Fig. 1a. A junction with shunt inductance can have multiple stable operating points (20) if $\beta = 2\pi I_0^c (L_s + L_M + L)/2\Phi_0$ >1; here $L_s + L_M + L \approx 4.2$ nH and $I_{c0} \approx 1.6 \mu A$, giving $\beta \sim 20/2=10$. For these parameters, Supplementary Fig. 1b shows the expected behavior of the internal flux in the loop $\Phi$ versus external current bias $I_{cb}$, where the stable operating points are branches with positive slope A, B, C, D and E (solid lines), and unstable operating points are given by dashed lines. When the coupler is set to the bias labeled $I_{cb}^{ON}$, it therefore can assume (randomly) any of the flux values given by the intersection of the gray dotted line with the stable branches A-E.

The branch must be precisely reset to place the coupler to a known bias. This may be accomplished using a technique developed for junction-shunted phase qubits (22). For example, to place the coupler on branch C, the coupler bias is repeatedly varied between the two values $I_{cb}^+$ and $I_{cb}^-$. As occupation in any other branch will result in a switch out of that branch, with enough trials the coupler will eventually finds itself in C, the only stable branch. Supplementary Figure 1c shows how this reset protocol is simply integrated into the swap experiment.

To determine $I_{cb}^-$ and $I_{cb}^+$, a qubit can be used to detect when the coupler switches to a different branch. When switching happens, the direct-current coupling between the coupler and qubit causes a qubit bias shift, which in turn can be measured by monitoring the escape of the qubit $|0\rangle$ state. For this measurement, we must first map out the onset and completion of the escape of the $|0\rangle$ state as a function of coupler and qubit bias. In Supplementary Fig. 2a, we show the escape probability for the $|0\rangle$ state as a function of reset offset $I_{cb}^R = I_{cb}^- = I_{cb}^+$ and the bias of qubit A. We take the dotted line as the condition where the $|0\rangle$ state had not escaped, and the dashed line as when the $|0\rangle$ state had fully escaped. We then plot in the Supplementary Figs. 2b and 2c these conditions versus the two coupler biases $I_{cb}^+$ and $I_{cb}^-$. The data for the former case is shown in Supplementary Figure 2b, where the region enclosed by the dashed box gives the values of $I_{cb}^-$ and $I_{cb}^+$ where the coupler is properly reset, i.e. when the qubit does not escape from the $|0\rangle$ state due to changing of a branch. Supplementary Fig. 2c shows a check of this condition, as this same region should always have the $|0\rangle$ state fully escaped. For optimal reset, $I_{cb}^-$ and $I_{cb}^+$ were chosen at the center of the dashed box. Repeating these measurements for qubit B gave similar values.

Because a number of reset cycles must be used to reliably reset the coupler, the reset probability must be measured versus the number of resets. As shown in Supplementary Fig. 2d, we found that 30 cycles produced an acceptable error of $\sim 1.5 \times 10^{-4}$.

We next confirmed that $I_{cb}^-$ and $I_{cb}^+$ properly reset the coupler for all coupler and qubit bias values up to the critical current of the coupler junction. As shown in Supplementary Fig. 2e, we plot the escape probability of qubit A as a function of coupler and qubit bias. The slope in the curve is expected and is due to the direct current connection between the coupler and qubit as discussed below. The increase in slope at the two ends of the curve arises from the non-linearity of the qubit's Josephson inductance as the critical current is approached.

**Compensation for qubit bias shift due to coupler bias**

As mentioned above, the coupler and qubit biases are connected via a direct-current path. As a result, a bias applied to the coupler also shifts the qubit biases. This can be seen in Supplementary Figs. 3a and 3b, which show the pulse sequence and data for two-qubit spectroscopy as a function of coupler bias. As the coupler bias $I_{cb}$ increases, the qubit biases shift in a direction that increases the qubit resonance frequencies. The shifts are approximately the same for both qubits, and can be easily compensated for as shown in Supplementary Figs. 3c and 3d. Here, the pulse sequence and spectroscopy data are the same except the qubit biases also contain a compensation pulse. With compensation, the qubit frequencies are constant as a function of coupler bias, indicating the shifts have been minimized.

**Minimum resolution of coupling strength for swap experiment**

For low coupling strengths, energy relaxation ($T_1$) makes the swap oscillations decay before the occurrence of a full swap. This decay limits the measurement of the minimum coupling strength that can be resolved. To determine the minimum coupling strength, corresponding to the OFF data repeated in supplementary Fig. 4A, we performed simulations of a two-qubit coupled system that included the $T_1$~350 ns decay as measured for each qubit. We show in the Supplementary Figs. 4b-d simulations for coupling strengths of 0.1 MHz, 0.3 MHz, and 0.5 MHz. The best fit to the data occurs at a simulated coupling strength of 0.1 MHz, as for coupling strengths < 0.1 MHz the oscillations cannot be resolved due to $T_1$ decay. We take the minimum coupling strength to be bounded as no greater than 0.1 MHz, giving a lower bound on the on/off ratio of 100 MHz / 0.1 MHz = 1000.

# Figure 1

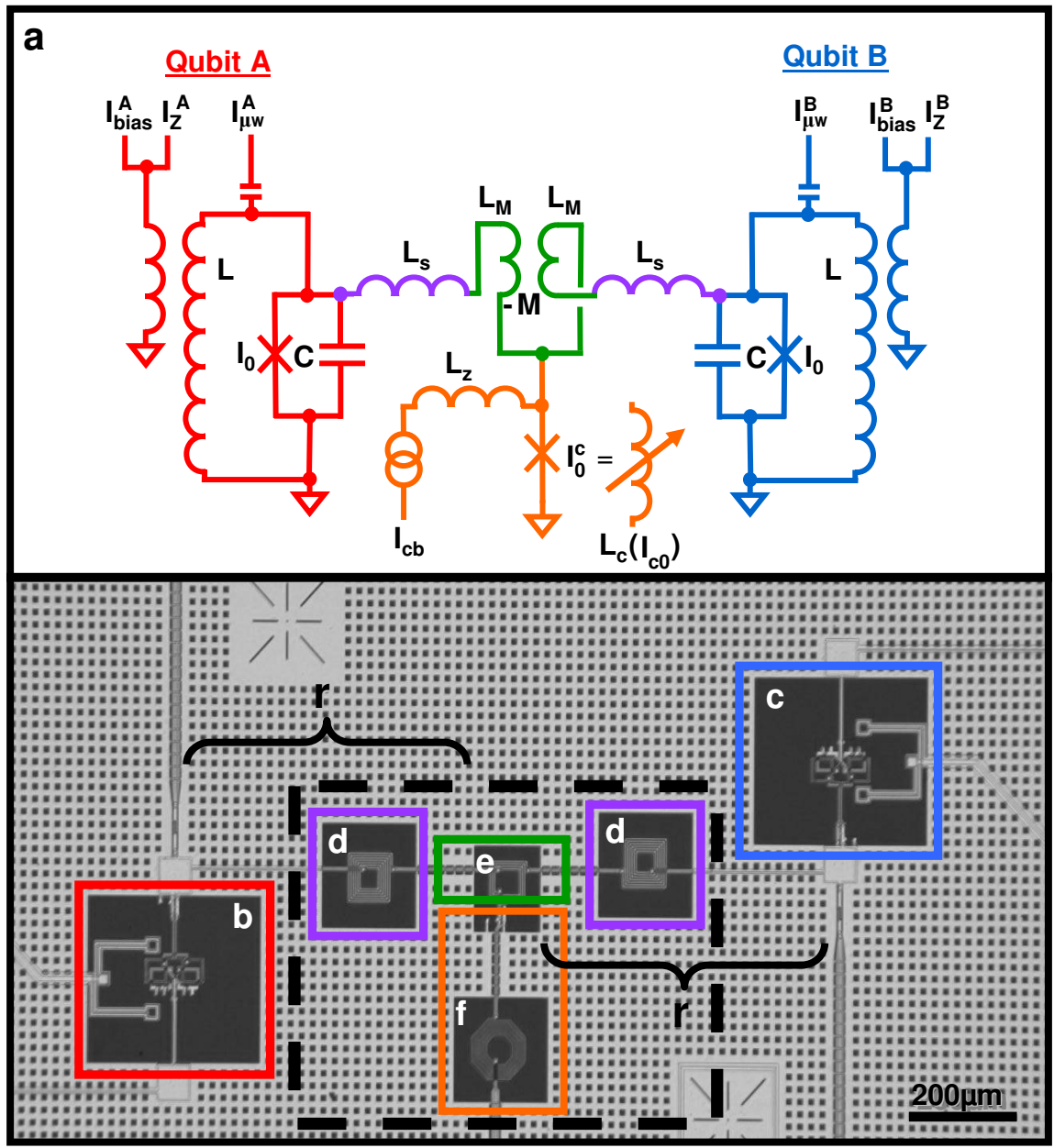

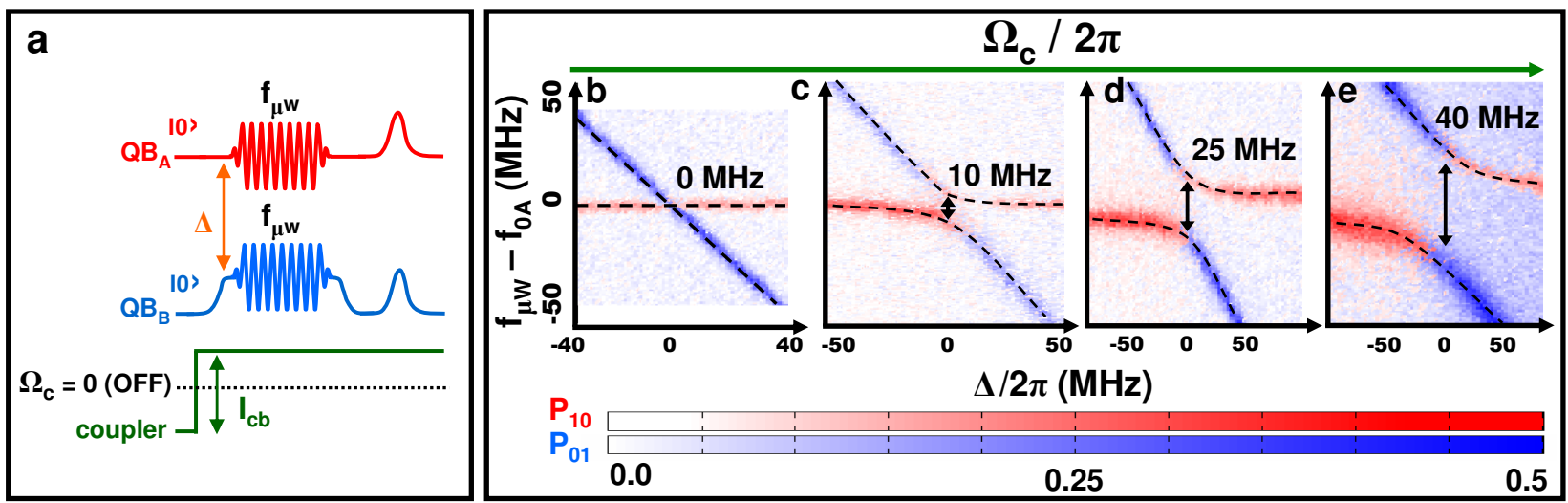

Figure 3

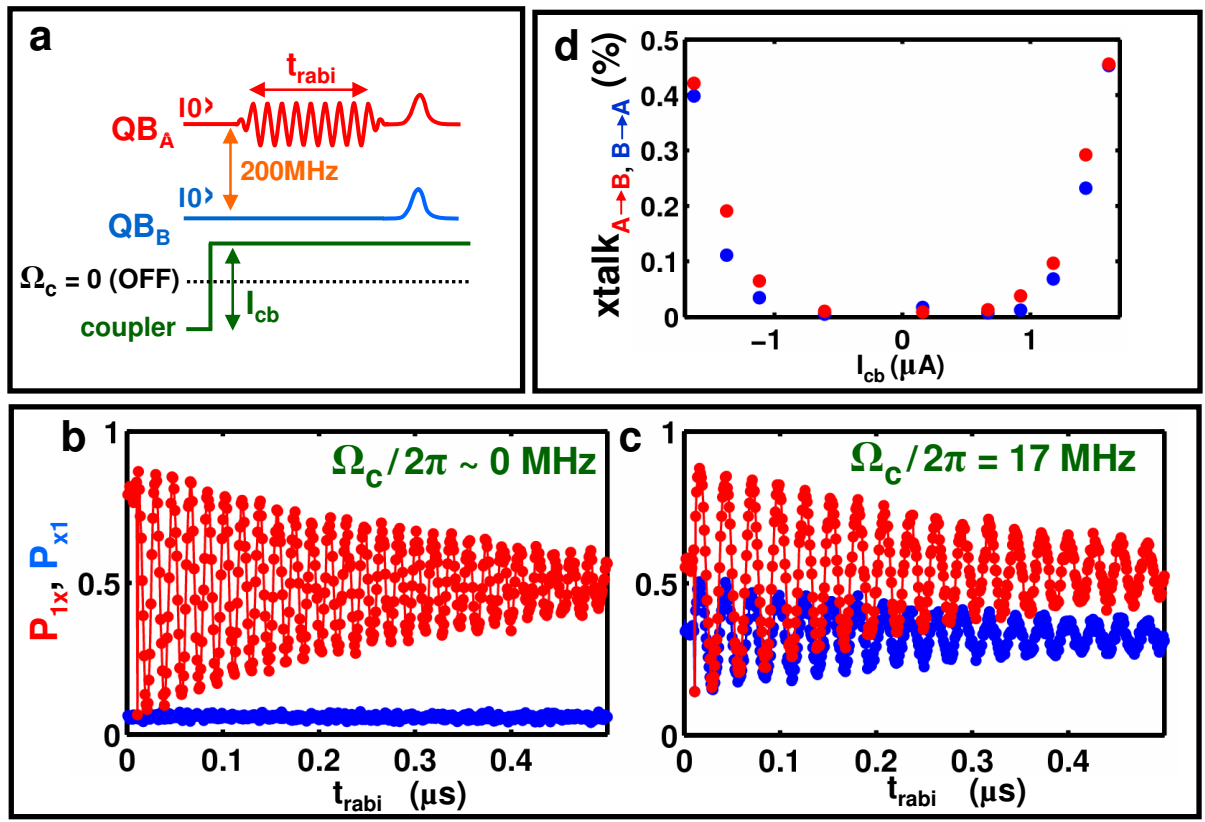

# Figure 4

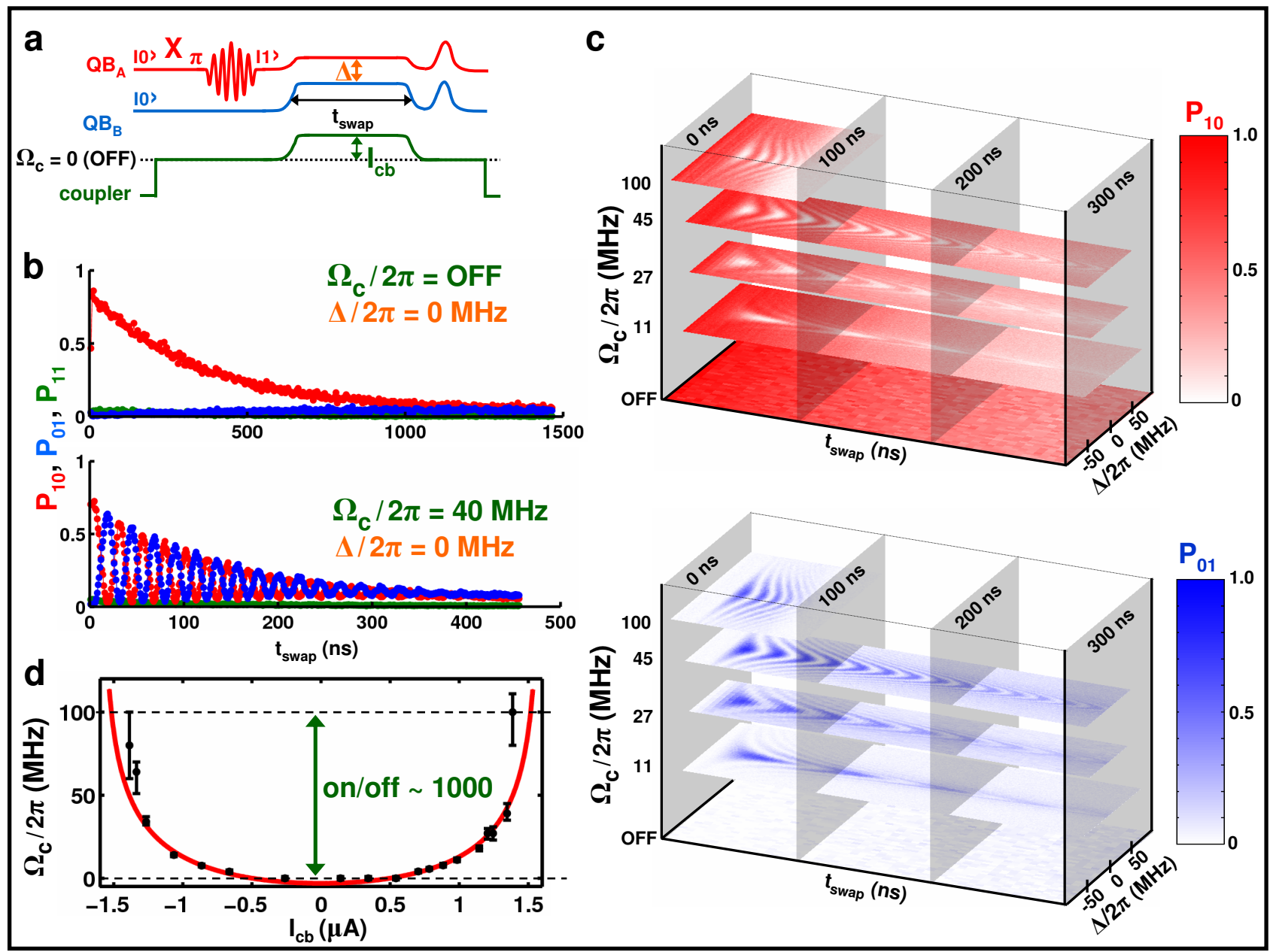

**Supplementary Figure 1**

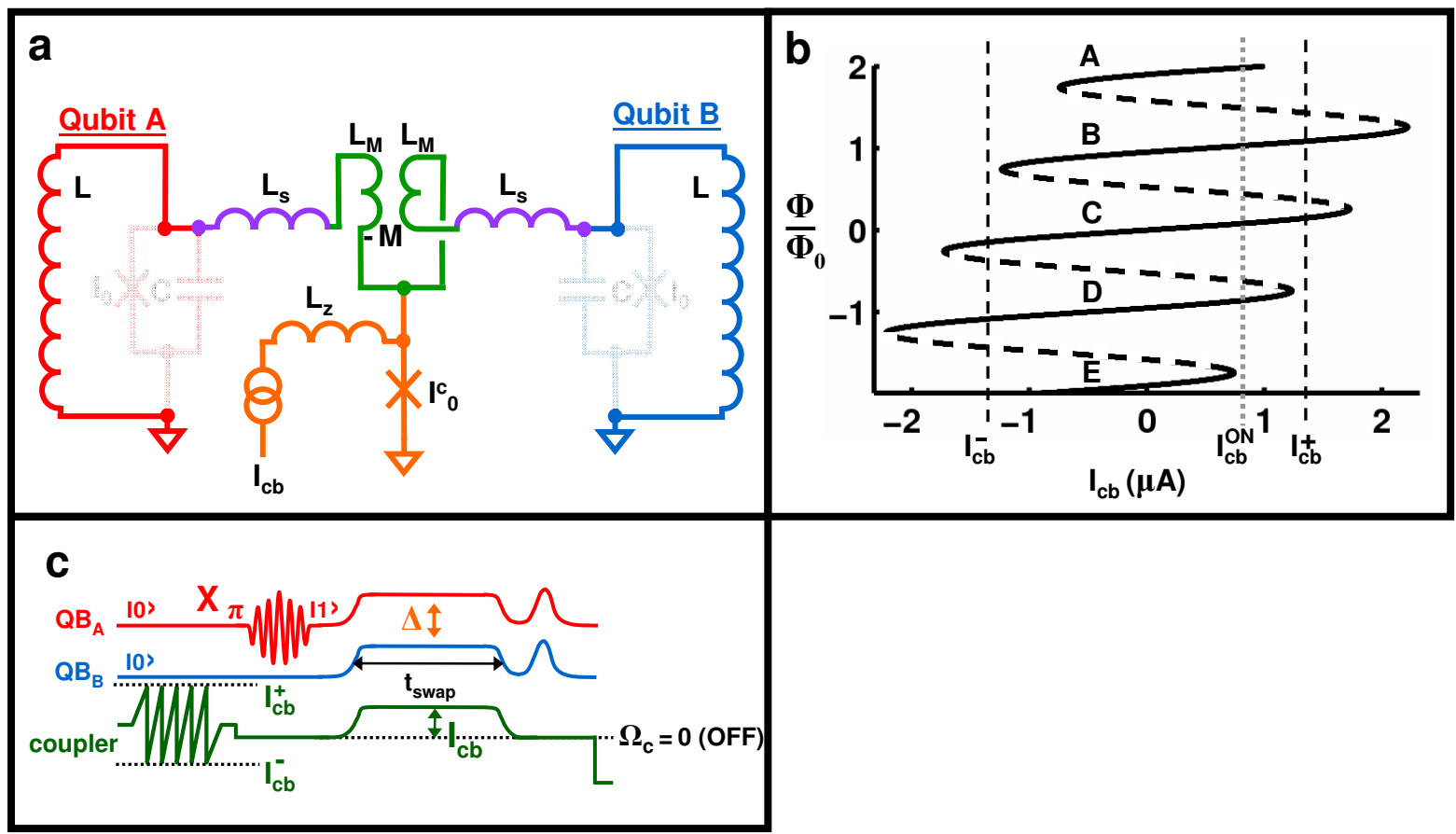

# Supplementary Figure 2

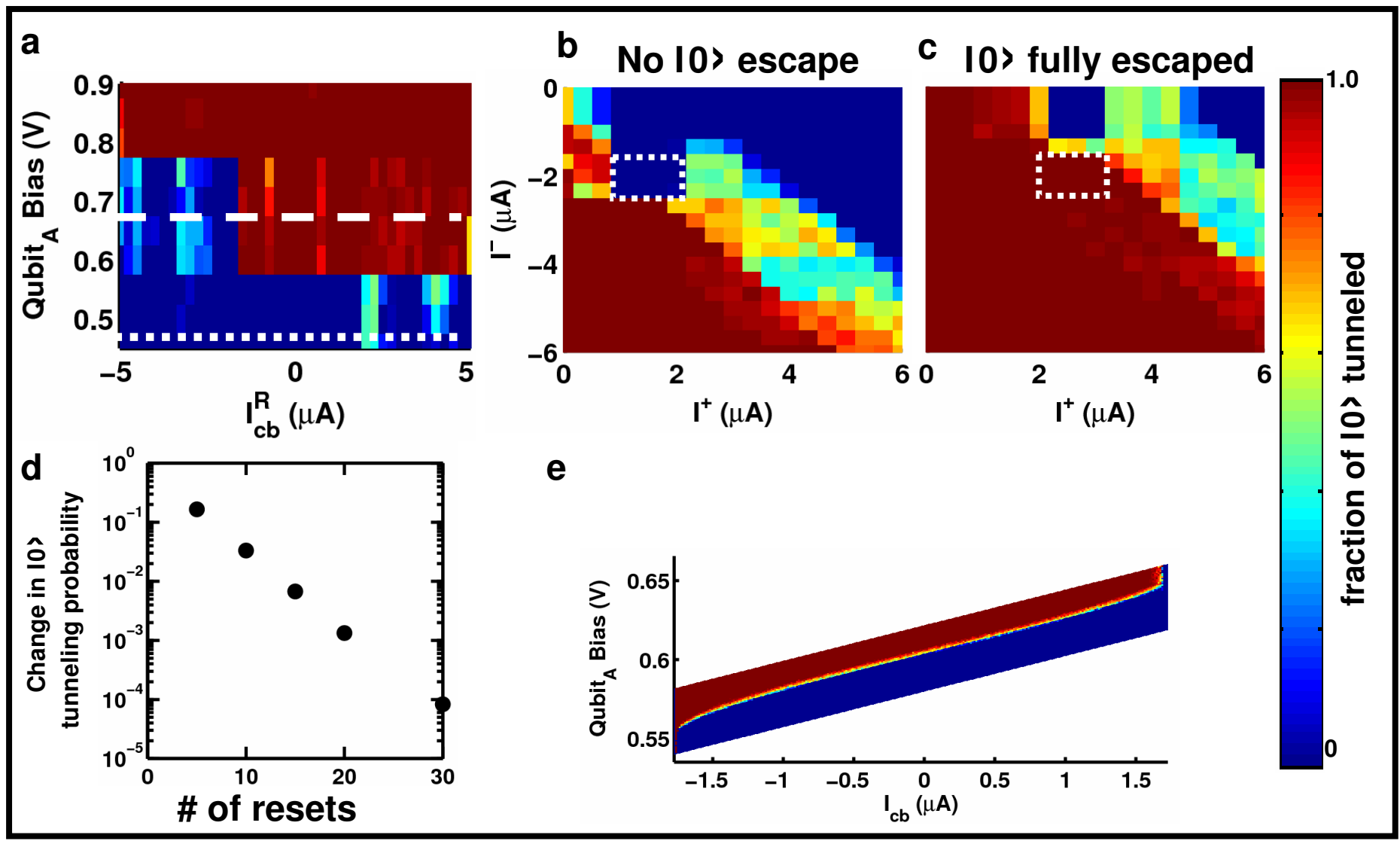

# Supplementary Figure 3

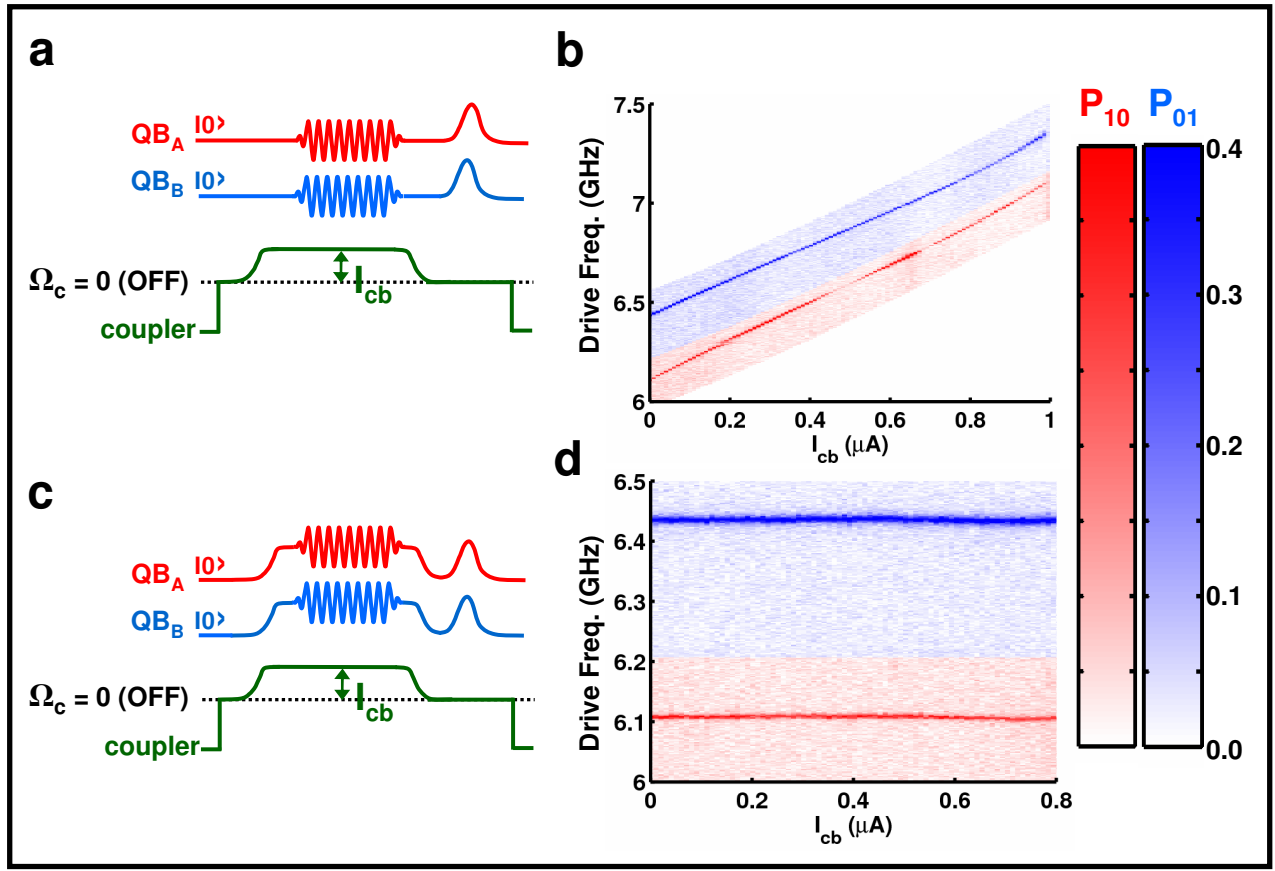

**Supplementary Figure 4**

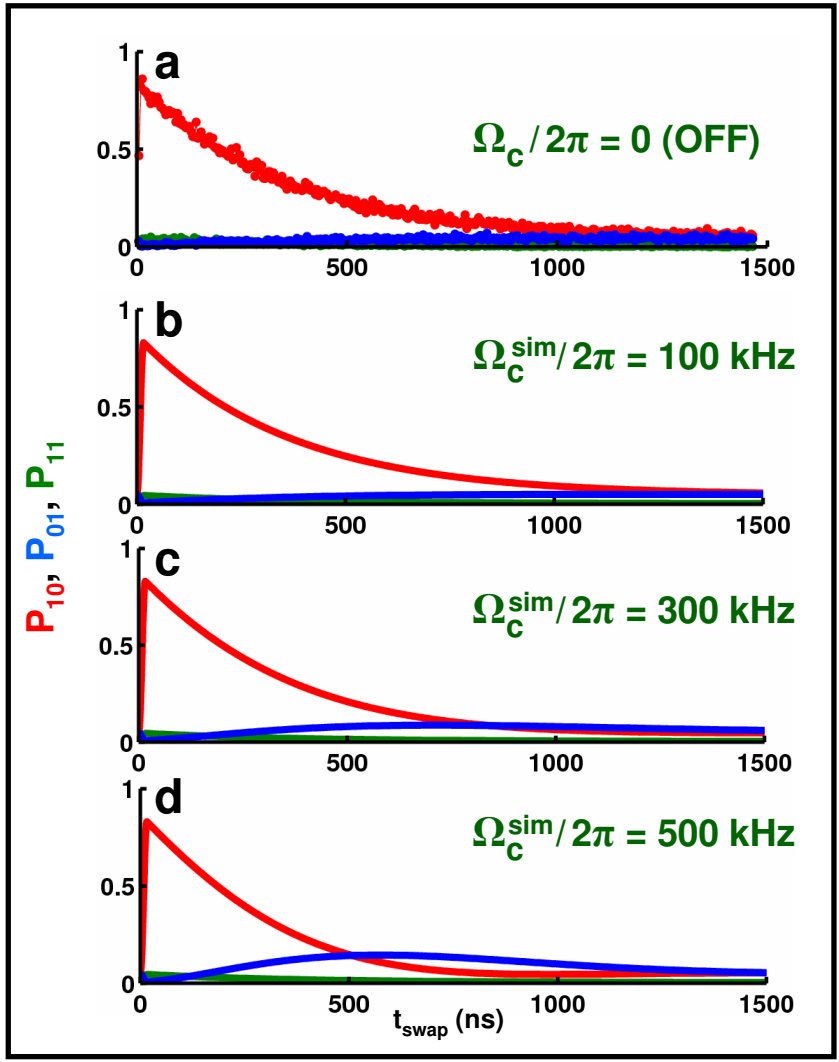